# Community Detection in Complex Networks by Dynamical Simplex Evolution

Vladimir Gudkov and Vladimir Montealegre

*Department of Physics and Astronomy, University of South Carolina, Columbia, SC 29208*

**Abstract.** We benchmark the dynamical simplex evolution (DSE) method with several of the currently available algorithms to detect communities in complex networks by comparing the fraction of correctly identified nodes for different levels of "fuzziness" of random networks composed of well defined communities. The potential benefits of the DSE method to detect hierarchical sub structures in complex networks are discussed.

The study of community detection in complex networks has become a subject of growing interest during recent years in a variety of areas. The reason for this interest is that partitioning a network into the groups of nodes that are more tightly linked (densely connected sets of nodes) is a crucial step to understand the structure, functionality and evolution of the whole network and its building constituents. This is useful for many practical purposes including the detection of real world network vulnerabilities. However, real world networks are usually very large, and community detection in complex networks is known to be a complete NP-problem [1], therefore, the computational demand required is very large, especially if a good level of accuracy is needed [2]. Many methods have been proposed to solve the problem efficiently, and their approaches include spectral analysis, hierarchical clustering methods; more recently a lot of attention has been drawn by the optimization of a quantity known as modularity [3-8].

Modularity is essentially a comparison measure of the number of links inside the detected modules of a network with the expected number of links that a random network with the same size and degree distribution would have. Community detection by modularity optimization uses the concept of betweenness to achieve a division algorithm that removes progressively the links with largest betweenness until the network breaks up into components. However, the properties of modularity have not been fully studied, and a clustering approach based upon its optimization has an intrinsic limit for the degree of resolution it can achieve (that depends on the number of links in the network) [9]. The existence of a resolution limit for community detection implies that it is *a priori* impossible to tell whether a module contains substructure (if smaller clusters can be refined inside it), and this cannot be overlooked if the network is one with a self similar nature (e.g. scale-free network), in which case a single partition does not describe the structure completely; therefore, for such a type of network a tree-like partition that digs into different structure levels is more appropriate.

In order to detect substructures inside the network, we require a method that does not discard links at any stage of the process, since every link contains information about the topology of the network. In this paper we use a dynamical simplex evolution (DSE) method [10] that has been shown to be effective for the purpose of detecting hierarchical structures in networks [11] and we compare it with the methods mentioned above. The main idea of the physics-based method is to eliminate *n!* permutations from the possible network description, in terms of the adjacency matrix, by choosing absolutely symmetric initial conditions: all *n* nodes are equidistant points (vertices) over a one (*n-1*) dimensional simplex and allowed to move as point-like objects. The forces acting on these nodes represent the links between nodes, and could be attractive or repulsive according to whether the pair of nodes are connected or disconnected. Therefore, the vertices have the tendency to move towards each other if there is a connection between them, and to repel if there is no connection. If there are only attraction forces, the vertices would collapse rapidly to a single spot that would represent the whole network. The introduction of repulsion forces reduces the speed at which the collapse happens, giving the vertices the opportunity to cluster correctly due to attractive forces while, at the same time, the repulsion forces provide a better separation of the groups of vertices that are less connected, helping in the identification of different clusters. A very important feature of the algorithm is the uniqueness of the solution. This is guaranteed by the choice of the forces: their magnitudes are independent of the relative distances between vertices, which leads to only one minimum of the multi-dimensional potential, and, as a consequence, to a single (unique) solution. To simplify calculations further and to avoid "overshooting" in the positions of vertices, the algorithm considers the motion of vertices as a motion of objects in a liquid with a large viscosity coefficient. This allows us to work with the first order differential equations instead of Newton's second order equations [10].

The dynamics of the vertices is governed by the forces (from the connectivity matrix) and the vertices displacements vary from vertex to vertex according to their mutual connectivities. Thus, after a small number of steps the new vertex positions depict accurately the cluster structure of the network, since the mutual distances between vertices which belong to the same cluster, are systematically smaller then the distances between vertices from different clusters. However, if the network's communities become fuzzier (if the number of links between clusters increases), a larger number of steps is required to separate clusters unambiguously. For instance, if the clusters are totally disconnected from each other (zero links between different communities), the attraction forces group the connected nodes as single sub networks, and the repulsion forces separate sharply the grouped sub networks from each other at every subsequent step, but if the number of links between clusters increases, the attraction forces between connected members in different clusters will link the sub networks as part of a larger community and despite of the presence of repulsion forces, the different clusters will eventually merge and collapse to a single spot (if the number of steps allowed is large enough). Thus the identification of the communities can be performed by choosing an adequate maximum threshold for the mutual distances (in the *n*-1 dimensional space), and then, collecting the nodes in groups that correspond to neighborhoods formed with vertices with mutual distances smaller than the chosen threshold. Since this method groups the vertices according to how tightly the nodes are connected to each other, finer substructure detection can be achieved by choosing a smaller threshold which provides a better resolution. At each step of a single run of the algorithm one can apply a set of different thresholds which provides instant (spectroscopic) resolution for the cluster and all sub-cluster structure of the network.

The maximum number of possible connections in a network is $M = n(n-1)$ where $n$ is the size of the network, however, the number of links in practical complex networks (real world and simulated) is usually only a fraction of the maximum number of possible connections $M$; this amplifies largely the contribution of the repulsive forces in the early steps of the vertex grouping process, which is very convenient to achieve a successful separation of the communities with only a few steps, however, if the number of connections is outnumbered by the number of disconnections (over the whole connectivity matrix) this becomes counterproductive, since a fast explosion of the vertices due to a large proportion of repulsion forces prevents them from gathering promptly into their communities before all the vertices collapse to a single spot. To thwart this effect, we seize the fact that the intensity of the forces can be chosen based on the density of the network connections and they can be considered as 'free' parameters of the algorithm. Thus we weigh the intensities of the attractive (repulsive) forces by using the ratio of the degrees of the nodes to the total number of nodes in the network by defining: $r_i = d_i / n$ where $d_i$ is the degree of node $i$ and $n$ is the total number of nodes in the network., Therefore, by using $I_{ij} = (1-r_i)C_{ij} - r_i(\neg C_{ij} - 1)$, we obtain a matrix for the intensities of the forces that already considers their signs. The forces $F_{ij}$ can then be calculated using the intensities $I_{ij}$.

It has become customary to test the efficiency of the clustering algorithms over a set of computer generated random networks with a well defined modular structure [12]. The benchmark networks have 128 nodes, a total of 1048 links, and they are composed of 4 clusters containing 32 nodes each, the nodes are connected with a probability $p_{in}$ ($p_{out}$) for members of the same community (different communities), in a way such that the average degree of every node in the network is $\langle k \rangle = 16$ (this provides control over the average number of links each node has with members of other communities $z_{out}$). As the number $z_{out}$ increases, the number of connections inside each cluster becomes smaller, so the structure becomes fuzzier and more difficult to identify, therefore, we can quantify the efficiency of the detection method by finding the fraction of correctly identified nodes $p$ for each value of $z_{out}$. However, there is no formal definition of the meaning of correctly identified nodes, therefore, we will use the following procedure to calculate $p$: Let $\{C_{i=1,2,3,4}\}$ be a class composed of the sets of nodes defined as clusters 1, 2, 3, 4 in the generated random network and $\{D_{j=1,...,m}\}$ be the class composed of the clusters detected by the algorithm, then, define $M_{ij} = |C_i \cap D_j|$ to be the number of nodes shared by sets $C_i$ and $D_j$, and we identify the cluster $C_i$ in the class of identified clusters to be $D_{k(i)}$ where $k(i)$ refers to the index of the set in class $D$ that maximizes the number of shared nodes. The value of $p$ can then be obtained by $p = \left(\sum_i M_{ik(i)}\right)/128$.

To illustrate how we can find an adequate value for the mutual distance threshold, we consider a random network generated by the procedure mentioned above with an average number of links per node between communities $z_{out} = 4.92$. We shuffle the locations of the nodes in the connectivity matrix of the network and let the BP algorithm run for only six steps. Figure 1 shows the histograms of the mutual distances for every pair of nodes in the network (fig. 1a), for connected nodes only (fig. 1b) and for disconnected nodes only (fig. 1c). Two humps are clearly visible in the three histograms, and in each case, the left hand hump corresponds to the mutual distances between nodes inside the same community while the right hand hump corresponds to the mutual distances between nodes belonging to different communities. Because of the symmetry of the matrix (the clusters have exactly the same sizes and the same densities) we see only two of these humps, but if the symmetry is broken we should expect to see a superposition of humps, each corresponding to a different cluster with a different density and size.

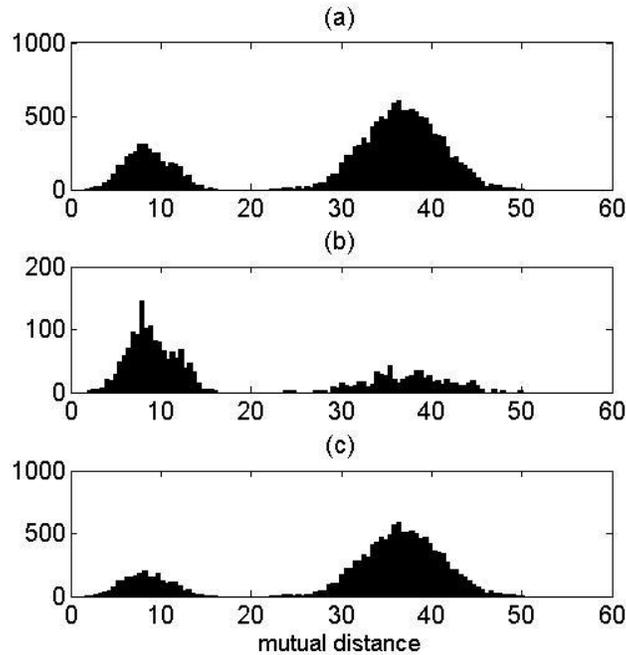

**FIGURE 1.** Histograms for the mutual distances of (a) every two nodes (b) only connected nodes and (c) only disconnected nodes in a generated random matrix with 128 nodes and four clusters, each containing 32 nodes.

This demonstrates that the structure can be resolved successfully. Fig. 1a shows that the maximum distance between nodes belonging to the same community is approximately 17 (in renormalized units for mutual distances), thus, by setting the maximum distance threshold in the surroundings of about 20, one can expect a good identification of the communities. The calculations for this network yield a fraction of correctly identified nodes of $p = 0.9922$ with a threshold of 15.4528. For the case of a network composed of substructures with different sizes and topologies, the histogram would not be as simple but rather would reflect the internal structure of all clusters and sub-clusters. Then the hierarchical structures can be identified by means of filtering the mutual distances between connected nodes and by grouping the nodes within adequate ranges of the filtered mutual distances.

The benchmark test for the DSE clustering method yields excellent results as shown in fig. 2. This figure presents the results of the algorithm test given in the paper [5]. Our results correspond to the line with stars; all other lines correspond exactly the figure 1 of the paper [5]. (For detailed discussions of the algorithms see [5-8] and references therein). The value of $p$, corresponding to each value of $z_{out}$ for the DSE algorithm, is obtained after averaging over 50 runs, (each run is the sequence of 30 steps).

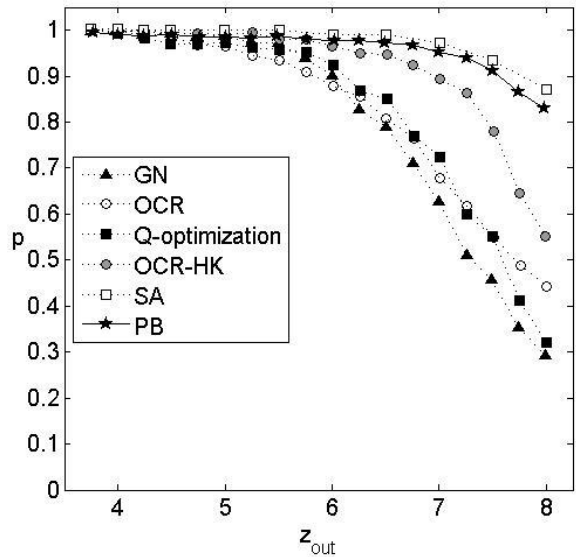

**FIGURE 2.** Fraction $p$ of correctly identified nodes as a function of $z_{out}$ (average number of links between clusters per node) for computer generated random graphs with 128 nodes.

It can be seen that the values of the fraction of correctly identified nodes for the DSE algorithm are very close to the values obtained with the SA (Simulated Annealing) model, their difference is very small and is only noticeable for $z_{out}>7$. Considering the fact that the SA model can be very demanding for computer time [13], it is quite remarkable that the DSE algorithm can achieve such good results with a number of $O(n^2)$ operations. The DSE algorithm is therefore very efficient in classifying the nodes, and at $z_{out}=8$ the average fraction of correctly identified nodes is $p>0.8$, which is a great improvement compared to the OCR methods [5] ($p>0.4$ for OCR and $p>0.5$ for OCR-HK). It should be noted, that the DSE algorithm provides additional detailed spectroscopic information about internal structures of sub-clusters in the same run of the algorithm automatically, which is an excellent feature for analysis of networks.

One can conclude that the DSE method possesses a variety of appealing properties considering the next interests. First, it does not require discarding (or weakening) connections progressively until a partition is observed. This allows the nodes to "interact" naturally, and based on complete information of the network at every stage of the algorithm, avoiding the possibility of biasing the results by modifying the network structure. Secondly, it permits the identification of substructures at different scales by setting adequate thresholds in the mutual distances for every step, which, at the same time, adjust the resolution. For example, for fuzzy structured networks (when the number of connections between communities is large), the resolution in cluster identification can be enhanced by increasing the number of steps and/or decreasing the step sizes. This is extremely important when the network has a self-similar nature, and hierarchical clustering is therefore expected. Another benefit of the BP method is the fact that it requires $O(n^2)$ operations, which is a great advantage for the CPU time required by it.

# ACKNOWLEDGMENTS

This work was supported by DARPA through AFRL grant FA8750-04-2-0260.